\documentclass[graybox]{svmult}

\usepackage{type1cm}        
\usepackage{makeidx}         
\usepackage{graphicx}        
\usepackage{multicol}        
\usepackage[bottom]{footmisc}

\usepackage{newtxtext}       %
\usepackage[varvw]{newtxmath}       
\usepackage{cite}

\newcommand{\ve}[1][K]{\textit{\textbf{#1}}}

\makeindex             
                       
\begin{document}

\title*{Target search kinetics for random walkers with memory}
\author{Olivier Bénichou, Thomas Guérin, Nicolas Levernier, Raphaël Voituriez}
\institute{Olivier Bénichou \at LPTMC, CNRS / Sorbonne Université, 4, Place Jussieu, 75252 Paris Cedex 05, FRANCE
  \email{olivier.benichou@sorbonne-universite.fr}
\and Thomas Guérin \at LOMA, CNRS / University of Bordeaux, 351 cours de la libération, 33400 TALENCE, FRANCE \email{thomas.guerin@u-bordeaux.fr}
\and Nicolas Levernier \at CINaM, CNRS / Aix Marseille Univ, Marseille, FRANCE \email{nicolas.levernier@univ-amu.fr}
\and Raphaël Voituriez  \at LJP, CNRS / Sorbonne University, CNRS / Sorbonne Université, place Jussieu, 75252 Paris Cedex 05, FRANCE \email{raphael.voituriez@upmc.fr}
}
 
 \maketitle

\abstract*{SAME ABSTRACT}

\abstract{In this chapter, we consider the problem of a non-Markovian random walker (displaying memory effects) searching for a target. We review an approach that links the first passage statistics to the properties of trajectories followed by the random walker in the future of the first passage time. This approach holds in one and higher spatial dimensions, when the dynamics in the vicinity of the target is Gaussian, and it is applied to  three paradigmatic target search problems: the search for a target in confinement, the search for a rarely reached configuration (rare event kinetics), or the search for a target in infinite space, for processes featuring stationary increments or transient aging. The theory gives access to the mean first passage time (when it exists) or to the behavior of the survival probability at long times, and agrees with the available exact results obtained perturbatively for  examples of weakly non-Markovian processes. This general approach reveals that the characterization of the non-equilibrium state of the system at the instant of first passage is key to derive first-passage kinetics, and  provides a new methodology, via the analysis of trajectories after the first-passage, to make it quantitative. 
 }

\section{Introduction}

How much time does it take for a random walker to go from a point $A$ to a point $B$, or to meet another particle  ? Answering this question amounts to solving a first passage problem \cite{Redner:2001a}:  one has to calculate the first passage time $T$ to reach a ``target region''. 
First passage problems appear naturally in many areas of physics, ranging from transport controlled reactions \cite{benAvraham2000,Berg1985}  (since a reaction cannot be faster than the time required for the reactants to come into contact), finance \cite{chicheportiche2014some}, polymer translocation \cite{palyulin2014polymer}, or search problems in biophysics (a virus searching for the nucleus of a eukaryotic cell \cite{Dinh2005}, a transcription factor searching for a gene sequence \cite{coppey2004kinetics}, etc). 
In this chapter, we  consider  the generic picture of a   random walker of position $\ve[r](t)$ evolving with time $t$ in a continuous $d$-dimensional space.  

In the last decades, the study of first passage problems has attracted a lot of attention in the literature \cite{Redner:2001a,metzler2014first,ReviewBray,benichou2014first}. Historically,  results have been derived first for classical random walks (or Brownian motion) on regular lattices (or uniform Euclidean spaces), and then extended to complex media  for processes that are memoryless (Markovian), for which the properties of the random walk at time $t+dt$ can be deduced from the knowledge of the position of the random walker only. For such processes, exact analytical tools, such as the renewal equation or backward Fokker-Planck equations, are available as starting points to study first passage properties~\cite{Redner:2001a,VanKampen1992,gardiner1983handbook}. However, when a random walker interacts with other variables of a "bath" (which could be internal degrees of freedom, or variables associated to the environment), the position of the random walker only must be described as a non-Markovian process, where memory effects appear when the dynamics of the other variables that interact with the random walker is ignored. A simple example is given by the dynamics of a tagged monomer of a polymer chain.  In this case  the position of the tagged monomer $\ve[r](t+dt)$ at $t+dt$ does not depend on $\ve[r](t)$ only but also on the positions of all other monomers of the polymer. This collective dynamics makes the  motion of the tagged monomer non-Markovian \cite{Panja2010,Panja2010a,bullerjahn2011monomer}.

Existing theoretical approaches to first passage kinetics of non-Markovian processes can be schematically classified as follows. 
First,  for a few  examples of specific non-Markovian stochastic processes analytical solutions can be obtained. This is the case of the random acceleration process \cite{Bicout2000} and of processes generated by telegraphic noises \cite{masoliver1986first,hanggi1985first,masoliver1986firstBOUND,masoliver1986firstFREE}. Second, a standard approach was introduced in the field of reactions involving polymers by Wilemski and Fixman \cite{WILEMSKI1974a,WILEMSKI1974b}, and proved to be applicable outside the field of polymer dynamics. This approach consists in considering that the  degrees of freedom of the bath are always in a stationary state and will be referred to as a \textit{pseudo-Markovian approach}. This approach has been improved in Ref.~\cite{Sokolov2003}. Even if useful \cite{kappler2019cyclization,Dua2002a,Campos2012,Hyeon2006,Debnath2004}, we will see that this approach can predict wrong scaling laws in some cases (see \text{e.g.} Ref.~\cite{Sanders2012} and below).  A third approach consists in developing a perturbation theory for weakly non-Markovian processes \cite{Wiese2011,delorme2015maximum,delorme2016perturbative,delorme2017pickands,sadhu2018generalized,wiese2019first,arutkin2020extreme}, assumed to be "close" to Brownian motion. These exact results have been so far restricted to one-dimensional processes and necessitate additional hypotheses  to be extended to the non-perturbative regime.  Last, a lot of work has been performed  for the problem of first passage in infinite space for non-stationary initial conditions, leading to the determination of persistence exponents characterizing the long-time decay of the first passage distribution (see review in \cite{ReviewBray}). 

\begin{figure}
\includegraphics[width=\linewidth]{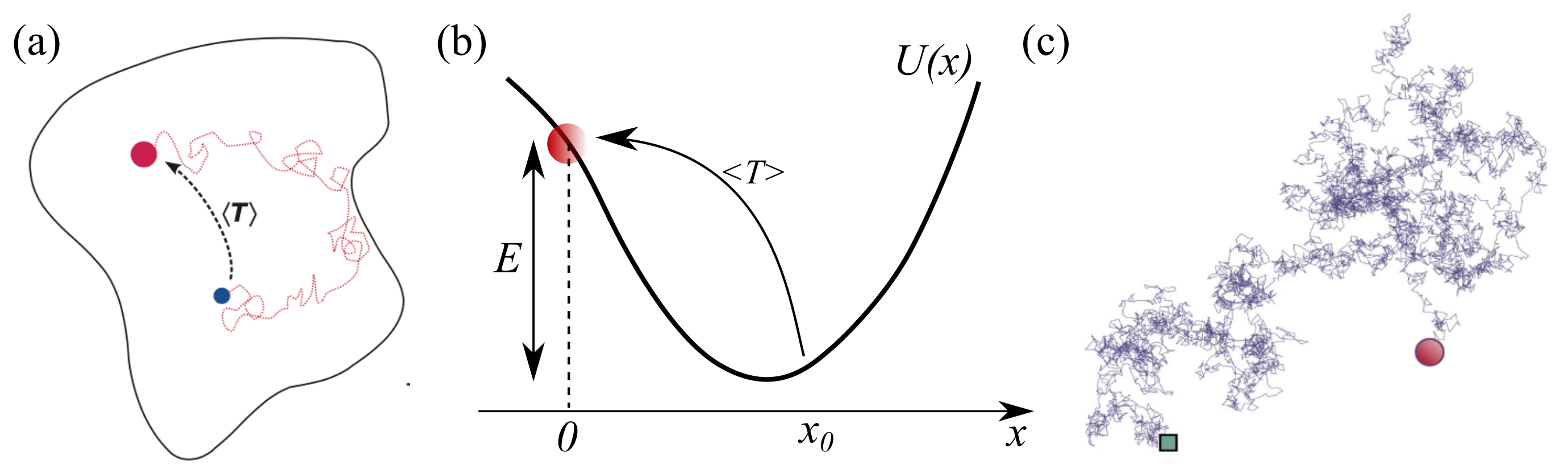}   
\caption{The three first passage problems that will be investigated in this chapter. (a) Search for a target in confinement. (b) Search, when $x(t)$ is a reaction coordinate, for an energetically costly configuration (rare event kinetics). (c) Search for a target in infinite space. 
} 
\label{ParadigmFPT}
\end{figure}

The goal of this chapter is to review recent works that introduced  an alternative  non-perturbative theoretical approach. This relies on the analysis of the trajectories followed by the random walker after the first-passage event, and can be developed in any space dimension~\cite{Levernier2022Everlasting,levernier2020kinetics,levernier2019survival,guerin2016mean}. We will show that this approach provides quantitative results for first passage properties in the three paradigmatic cases  represented in Figure \ref{ParadigmFPT}. 
\begin{enumerate}
\item{ In the first situation, shown on Fig \ref{ParadigmFPT}(a), a random walker is looking for a target in a confined domain. In this case the random walker is able to explore the whole volume of the domain, and there is no substantial energy barrier to cross to reach the target. Such geometric confinement can be achieved by hard walls or softer confinement  (such as harmonic trapping).  The target search can be seen as \textit{limited by entropy} since what limits the search is the exploration of the volume. }
\item{In the second case, the random walker is assumed to describe a reaction coordinate $x(t)$ moving in an energy landscape $U(x)$ and the target is located at a configuration of high energy [Fig \ref{ParadigmFPT}(b)]. The first passage is in this case a \textit{rare event} and is limited by the energy cost to reach the target.}
\item{In the third case of exploration in infinite space [Fig \ref{ParadigmFPT}(c)], where the stationary probability density in the absence of target vanishes $p_s(\ve[r])=0$, first passage properties are quantified by the survival probability, which is characterized  by a persistent exponent defining its algebraic decay, and a prefactor. These will be discussed  for processes with stationary increments, and for processes with  non-stationary initial conditions.}
\end{enumerate}
The presentation is based on results of Refs.~\cite{Levernier2022Everlasting,levernier2020kinetics,levernier2019survival,guerin2016mean}.
Note that   we will leave  aside the case of first passage properties for anomalous transport coming from jumps with broadly distributed waiting times and/or jump lengths, such as in Continuous Time Random Walks or Lévy walks. Strictly speaking, such random walks are also non-Markovian if the waiting times are not exponentially distributed. However, such processes are amenable to Markovian analysis, essentially because  renewal equations can be obtained \cite{hughes1995random,ReviewBray,Meyer2011,levernier2018universal}. 

The outline of this chapter is as follows. First, we derive a general equation for the mean first passage time, for non-Markovian non-smooth random processes (Section \ref{GenForm}). Then, we show how to predict the mean first passage time for non-Markovian random walkers searching for a target  in a large volume (Section \ref{SectionTargetSearch}). Next, we show how the formalism can be adapted to describe the kinetics of rare events in Section \ref{RESection}. Next, we describe the first passage without confinement and describe how to obtain the long-time asymptotics for the survival probability for processes with stationary increments (Section \ref{SurvProbSection}). Finally, in Section \ref{PersistenceSec}, we show how the formalism can be adapted to calculate persistence exponents in the case of  non stationary initial conditions, such as after a quench.

\section{Mean first passage times for processes reaching a stationary state}
\label{SectionMFPT}

\subsection{General formula for the mean FPT for a non-Markovian, non-smooth random walker}
\label{GenForm}

We start this section with the derivation of a general relation for the mean first passage time for a non-Markovian random walker $x(t)$ in one dimension, with the hypothesis that $x(t)$ is continuous, non-smooth (so that $\langle \dot{x}^2\rangle=\infty$ \cite{ReviewBray}, as for overdamped processes), and reaches a stationary state at long time. 
We denote by $p(x,t)$ its probability density function (PDF) at time $t$, and the stationary PDF is $p_s(x)=\lim_{t\to\infty} p(x,t)$. Both $p$ and $p_s$ are defined in the absence of a target. Now, we call $T$ the first passage time (FPT) to a target at $x=0$, and $F(T)$ the PDF of FPTs.  Our starting point is a ``tautological'' equation which comes from the following remark: since the process is non-smooth, if the particle is observed at $x$ at $t$ then it means that the particle had already reached the target at some earlier time $\tau$ for the first time, with $\tau<t$. Then, we can write,
\begin{align}
p(0,t)=\int_0^t d\tau F(\tau)p(0,t\vert \text{FPT}=\tau) \label{Renewal},
\end{align}
where $p(0,t)$ is the probability density to be at $x=0$ (for the dynamics without absorption to the target) and $p(0,t\vert \text{FPT}=\tau)$ is the probability density to observe the particle at  $x=0$ at time $t$ given that  the FPT is $\tau$ (when the particle is allowed to continue its motion after the FPT). This equation is a generalized  ``renewal'' equation \cite{VanKampen1992}. To the best of our knowledge the fact that one should condition the propagators to the value of the FPT was first noted in Ref.~\cite{Likthman2006}. Here we note that in the Markovian case  one could write  $p(0,t\vert \text{FPT}=\tau)=p(0,t-\tau\vert0)$ so that the above equation would  involve a convolution, enabling one to derive a general expression of the FPT density and its moments (see eg \cite{Condamin2007,benichou2014first}). This is however not possible in the non-Markovian case where the convolution structure is lost and one sees that one has to characterize the probability density conditioned to the fact that a first passage event was observed.  

\begin{figure}[t!]
\centering
\includegraphics[width=\linewidth,clip]{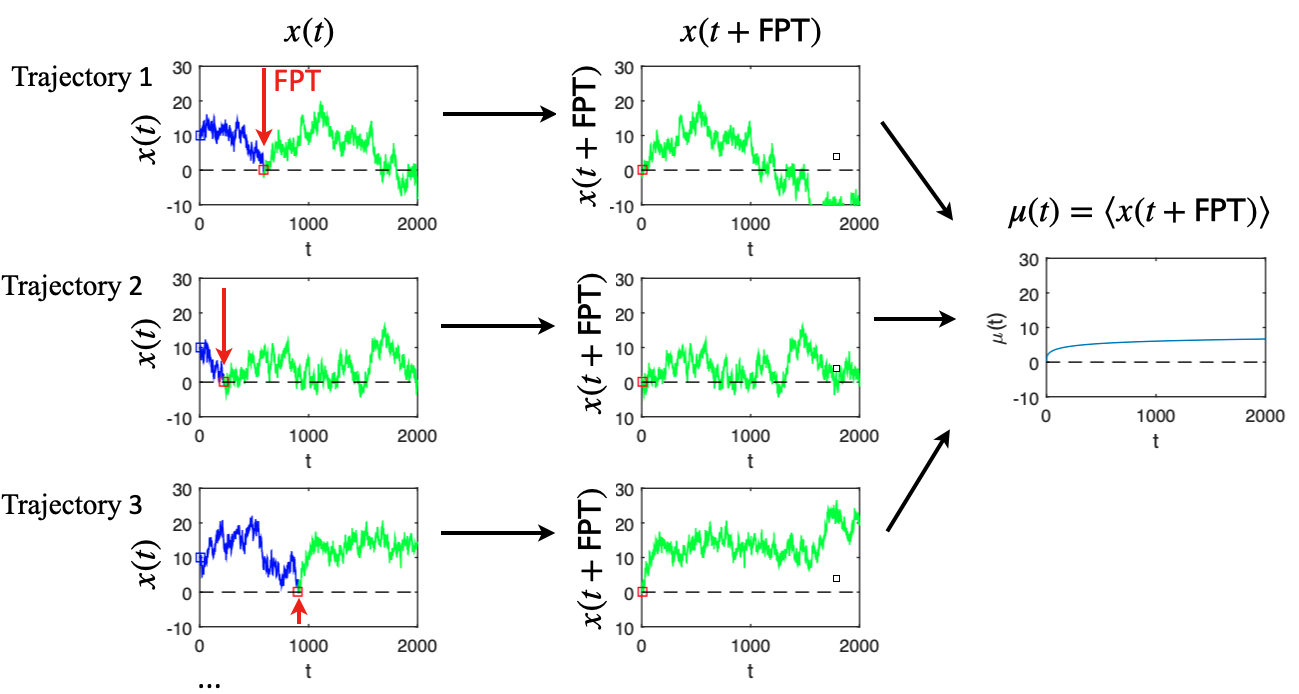}   
\caption{Construction of the average trajectories in the future of the FPT $\mu(t)$. For each realization of a stochastic trajectory, one measures the FPT (left) and uses this value of time as an initial time for the process in the future of the FPT.  Averaging those new trajectories over a large number of realizations, one obtains $\mu(t)$. Here, the trajectories are real examples of a scale invariant subdiffusive process of MSD $\psi(t)=t^{0.72}$.}
\label{FigureConstructionTrajectoires}
\end{figure}

Now, we introduce the process in the future of the FPT, $y_\pi(t)\equiv x(t+\text{FPT})$.  The process $y_\pi(t)$ is thus defined from a given stochastic trajectory $x(t)$ by choosing the first passage event as the initial condition, see Fig.~\ref{FigureConstructionTrajectoires}. The suffix $\pi$ is used for historical reasons as $\pi$ is sometimes used to denote the probability with which a part of an extended target is reached for the first time~\cite{VanKampen1992,Condamin2008}, and here $y_\pi$ is linked to the configuration of non-reactive degrees of freedom at the first passage~\cite{Guerin2012a}. We define   $p_\pi(y,t)$  as the probability density of observing $y_\pi(t)=y$, which reads 
\begin{align}
p_\pi(y,t )=\int_0^\infty d\tau F(\tau)p(y,t+\tau \vert \text{FPT}=\tau) \label{Def2PointsPPi}.
\end{align}
 Now we show how to make  the mean FPT appear by a few algebraic manipulations. We consider Eq.~(\ref{Renewal}), where we substract $p_s(0)$ on both sides, leading to
\begin{align}
p(0,t)-p_s(0) = \int_0^t d\tau F(\tau) [p(0,t \vert \text{FPT}=\tau)-p_s(0)]  - \int_t^\infty d\tau F(\tau)p_s(0)  \label{04381},
\end{align}
where we have used the fact that $\int_0^\infty d\tau F(\tau)=1$. Next, we write 
\begin{align}
\int_0^\infty dt \int_t^\infty d\tau F(\tau)  
= \langle T\rangle \label{Trick1},
\end{align}
which is obtained by changing the order of integration between $t$ and $\tau$. We also note the following equalities:
\begin{align}
 \int_0^\infty dt  \int_0^t d\tau F(\tau)&  [p(0,t \vert \text{FPT}=\tau)-p_s(0)] \nonumber\\
&=\int_0^\infty d\tau \int_{0}^\infty du \ F(\tau) \ [p(0,\tau+u\vert  \text{FPT}=\tau)-p_s(0)] \label{line2}\\
&=\int_0^\infty du \ [p_\pi(0,u) -p_s(0)]\label{Trick2},
\end{align}
where we have changed the order of  between $(t,\tau)$ and used  $t=u+\tau$ to obtain  Eq.~(\ref{line2}), and we have used the definition (\ref{Def2PointsPPi}) to simplify the resulting integral. 
Using Eqs.~(\ref{Trick1}) and (\ref{Trick2}), we see  that integrating Eq.~(\ref{04381}) over $t$ leads to \cite{guerin2016mean}
\begin{align}
\langle T\rangle p_s(0) = \int_0^\infty dt \ [p_\pi(0,t)-p(0,t)]. \label{GenMFPT}
\end{align}
This equation is general and exact, as soon as $p_s$ exists, for any continuous non-smooth stochastic process, even non-Gaussian.  
However, at this stage, it is not explicit since we have no information on $p_\pi$. 

We therefore consider a  two-point generalized version of the renewal equation: 
\begin{align}
&p(0,t;x_1,t+t_1)=\int_0^t d\tau F(\tau) p(0,t;x_1,t+t_1\vert\text{FPT}=\tau). \label{Ren2Points}
\end{align}
Here, $p(0,t;x_1,t+t_1)$ is the joint PDF of  $x=0$ at time $t$ and $x=x_1$ at a later time $t+t_1$.  Here $x_1$ and $t_1$ are arbitrarily fixed   scalars, with $t_1>0$. Multiplying by $x_1$ and integrating over $x_1$ leads to \cite{guerin2016mean}
\begin{align}
\int_0^\infty dt [p_\pi(0,t)\mathbb{E}_\pi(y_\pi(t+t_1)\vert y_\pi(t)=0)-p(0,t)\mathbb{E}(x(t+t_1) \vert x(t)=0)] =\nonumber\\
  \langle T\rangle p_s(0)\mathbb{E}_s(x(t_1)\vert x(t)=0),\label{05942}
\end{align}
where the notation $\mathbb{E}(A\vert E)$ stands for the conditional average of $A$ given that the event $E$ is realized, 
$\mathbb{E}_\pi$, $\mathbb{E}_s$ and  $\mathbb{E}$ denote the average over the respective PDFs $p_\pi,p_s$ and $p$. This general equation will be useful to characterize the mean first passage time in two limiting cases: the search for a target for a symmetric random walk in a large volume (Section \ref{SectionTargetSearch}), and the search for a rarely reached event (Section \ref{RESection}).

\subsection{Target search in a large volume}
\label{SectionTargetSearch}

\subsubsection{Explicit form of the equations for the mean FPT}
We consider now the more specific case of target search in confinement, when the following property is satisfied: we assume that there is a limit in which $p_s(0)\to0$, which we call large volume limit, corresponding to boundaries going far away from the  target  and the initial position of the walker. Note that, in the case of a random walk in an harmonic potential, such a limit is reached when the stiffness of the potential vanishes. We assume that, in this large volume limit, the trajectories of $x(t)$ are not biased (no preferred direction), that $x(t)$ has stationary increments (no aging), and that $x(t)$ is a Gaussian process. Of note this Gaussian hypothesis  does not imply that correlations are weak;  instead we will see that our theory can deal with  processes for which the relaxation time of the increments are formally infinite (such as in the case of the fractional Brownian motion).  We also note that there are many physical examples of Gaussian non-Markovian processes, such as the dynamics of tracer particles moving in complex viscoelastic fluids \cite{mason1997particle,mason1995optical}, nematic fluids \cite{Turiv2013}, crowded narrow channels \cite{wei2000single}, or attached to polymers \cite{Panja2010,Panja2010a}. 

With these hypotheses, one can show that $x(t)$ is fully characterized as soon as one specifies the Mean Square Displacement (MSD) $\psi(t)=\langle [x(t+\tau)-x(\tau)]^2\rangle$, since the covariance at times $t,t'$ takes the form
\begin{equation}
\sigma(t,t')=\text{Cov}(x(t),x(t'))=\frac{1}{2}\left[\psi(\vert t-t'\vert) -\psi(t)-\psi(t') \right].\label{sigmaEq}
\end{equation} 
We assume that the MSD diverges at long times  as  $\psi(t)\simeq\kappa t^{2H}$ for  $t\rightarrow\infty$, with $0<H<1$, with $\kappa>0$ a transport coefficient. The motion at long times can  then be either subdiffusive ($H<1/2$) or superdiffusive ($H>1/2$), and in all cases the particle is not trapped in the vicinity of the initial point but is free to explore  space. Note that $\psi(t)$ is defined in the absence of confinement, so that the behavior $\psi(t)\propto t^{2H}$ is not contradictory with the fact that the ``real'' MSD must saturate at times for which the random walker reaches the confining boundaries. 

In the large volume limit, it is found that all terms in Eqs.~(\ref{GenMFPT}) and (\ref{05942}) have a well defined, non vanishing infinite space limit, except for the term $p_s(0)$. Next, we assume that the trajectories in the future of the first passage event have Gaussian statistics, with the same covariance $\sigma(t,t')$ as the initial process, and an average $\mu(t)=\langle x(t+\text{FPT})\rangle$. With these approximations, we can calculate the conditional averages entering (\ref{05942}) by using existing formulas for conditional averages for Gaussian processes \cite{Eaton1983}, leading to 
\begin{align}
\int_0^\infty  \frac{dt}{ \sqrt{\psi(t)}}\left\{\left[\mu(t+\tau)-\mu(t)K(t,\tau)\right]e^{-\frac{\mu(t)^2}{2\psi(t)}}-x_0\left[1-K(t,\tau)\right]e^{-\frac{x_0^2}{2\psi(t)}}\right\}=0\label{Eq2A},
\end{align}
with $K(t,\tau)=[\psi(t+\tau)+\psi(t)-\psi(\tau)]/[2\psi(t)]$, while the mean FPT reads
\begin{align}
\langle T\rangle p_s(0)=  \int_0^{\infty}dt \frac{1}{[2\pi\psi(t)]^{1/2}}\left[e^{-\mu(t)^2/(2\psi(t))}-e^{-x_0^2/(2\psi(t)) }\right] \label{Eq1A}.
\end{align}

Second, the long-time properties of $\mu(t)$ can be obtained. Analyzing Eq.~(\ref{Eq2A}) for large $\tau$ leads (after some lines of algebra) to the conclusion  that
\begin{align}
\mu(t)\simeq x_0 -  \frac{B}{t^{1-2H}}, \hspace{2cm} (t\rightarrow\infty),\label{BehaviorMu}
\end{align}
where $B$ is a constant, usually positive, that depends on the behavior of the process at all timescales. This expression clearly shows that for subdiffusive processes ($H<1/2$) the trajectories in the future of the FPT converge to $x_0$ at long times. This initial position is therefore never forgotten, underlying the strongly non-Markovian nature of the problem. For superdiffusive processes, on the  contrary, the trajectory in the future of the FPT typically crosses the target  and eventually escapes to $-\infty$.  For processes with finite memory, i.e. which become diffusive at long times, $\mu(t)$ reaches a non-vanishing constant. Next, using (\ref{BehaviorMu}), it is clear that the integral  (\ref{Eq1A}) exists, meaning that the formalism predicts that the mean FPT scales linearly with the volume $V=1/p_s(0)$ in the large volume limit. 
This may be compared with the Wilemski-Fixman closure approximation \cite{WILEMSKI1974a,WILEMSKI1974b}. This approximation was developed in the context of reactions involving polymers,  and consists in assuming that the hidden degrees of freedom (the non-reactive monomers in a polymer chain) follow an equilibrium distribution (conditioned to the fact that reactive monomers are in contact). In our framework, this approach  translates into the approximation $\mu(t)\simeq 0$, which would incorrectly predict that  the mean FPT in (\ref{Eq1A}) diverges for $H<1/3$. 

\begin{figure}[t!]
\centering
\includegraphics[width=1\linewidth,clip]{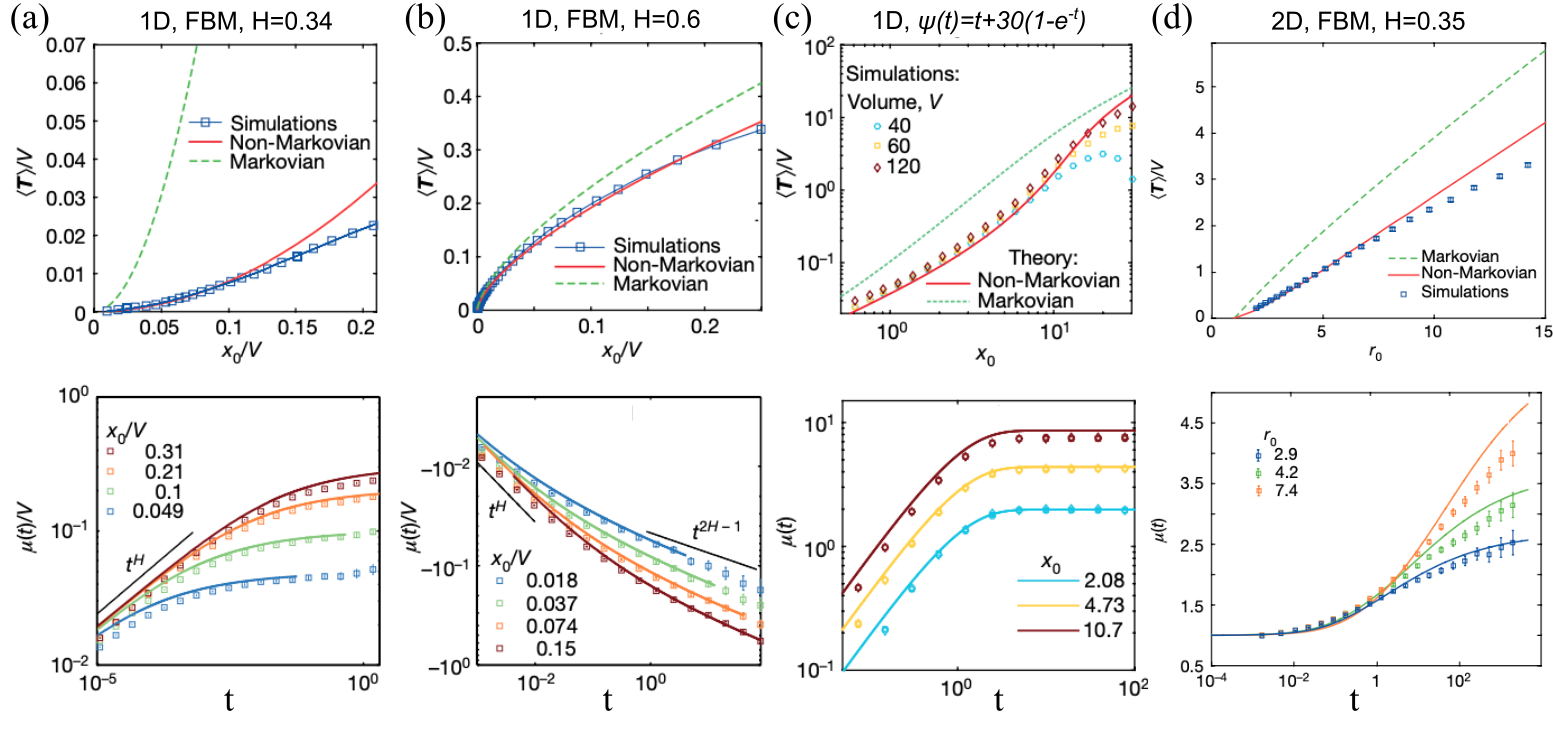}   
\caption{Test of the non-Markovian for several stochastic processes, upper graphs show the mean FPT, lower graphs display the value of $\mu(t)$. Symbols are the results of numerical simulations;  continuous lines are the results of the non-Markovian theory, green dashed lines are the results of the pseudo-Markovian approximation, with $\mu(t)\simeq0$. The value of the MSD are respectively   $\psi(t)=t^{2H}$ for (a),(b),(d) (with $H$ indicated in the graph) and $\psi(t)=30(1-e^{-t})+t$ for (c). The examples (a),(b),(c) are one-dimensional (1D), for (d) the problem is two-dimensional (2D), with a target radius $a=1$. Adapted from Ref.~\cite{guerin2016mean}. }
\label{FigTestTheory}
\end{figure}

Some predictions of this approach are presented on Fig.~\ref{FigTestTheory}, for various processes, which clearly show that it is much more precise than a pseudo-Markovian approximation. The examples of stochastic processes shown in these figures cover the case of diffusion in a Maxwell viscoelastic fluid  \cite{grimm2011brownian} and that of the fractional Brownian motion, where $\psi(t)$ is a power law, with infinite memory. 

\subsubsection{Case of scale invariant processes } 

Let us now consider the case where the process is scale invariant at all times (not only at long times): $\psi(t)=\kappa t^{2H}$. In this case we can show  that the solutions of (\ref{Eq2A}) satisfy: 
\begin{align}
\mu(t,x_0) = x_0 \ f_H\left( \frac{t \ \kappa^{1/2H}}{x_0^{1/H}} \right)  
\end{align}
where the function $f_H(u)$ depends only on $H$ and is the solution of (\ref{Eq2A}) in the particular case $\psi(t)=t^{2H}$ and $x_0=1$. Inserting this scaling into Eq.~(\ref{Eq1A})  leads to 
\begin{align}
&\langle T\rangle /V = \beta_H \frac{x_0^{1/H-1} }{\kappa^{1/2H}}, & \beta_H= \int_0^\infty du \frac{e^{-f_H^2(u)/2u^{2H}}-e^{-1/2u^{2H}} }{\sqrt{2\pi}u^H}.
\end{align}
This means that the scaling with the initial distance is the same as in the Markovian case \cite{Condamin2007},  with however a prefactor that takes into account non-Markovian effects encompassed in the function $f_H$. 

\subsubsection{Generalization to higher spatial dimensions  } 

All the arguments presented above can be adapted in $d$-dimensions. Assuming that the target is a (small) sphere of radius $a$, the problem is spherically symmetric in the large volume limit and it is useful to assume that the initial position is randomly distributed around the target, with a fixed initial distance $r_0$. Then the average trajectory in the future of the FPT can be written as $\mu(t)\hat{\ve[u]}(\theta)$, where $\theta$ is the angle at which the target is hit for the first time and $\hat{\ve[u]}$  the unit vector in the direction $\theta$. With these approximations, one obtains the self-consistent equations \cite{guerin2016mean}
\begin{align}
0=\int_{0}^{\infty} \frac{dt}{ \psi(t)^{d/2}}  &  \Bigg[    e^{-\mu(t)^{2}/[2\psi(t)]} \left( \frac{\mu(t)}{d \, \psi(t)} \left( \mu(t+\tau)-\mu(t)\frac{\sigma(t+\tau,t)}{\psi(t)} \right)  + M(t,\tau)\right)   \nonumber\\
  &    - e^{-r_{0}^{2}/[2\psi(t)]} \left( \frac{r_{0}}{d \, \psi(t)} \left( r_{0}-r_{0}\frac{\sigma(t+\tau,t)}{\psi(t)} \right)  +M(t,\tau) \right) \Bigg].
\end{align}
with $M(t,\tau)=[\sigma(t+\tau,t)-\sigma(t,t)]/\psi(t)$, and $r_0$ is the initial distance to the center of the target. This equation is valid for $d=2$ and $d=3$. Note that the radius of the target $a$ enters in the problem by imposing $\mu(t=0)=a$. The mean FPT to the target is  
\begin{align}
\frac{\langle T\rangle }{V}=  \int_{0}^{\infty} {dt}[p_\pi(\ve[0],t)-p(\ve[0],t)]= \int_{0}^{\infty} dt \frac{     e^{-\mu(t)^{2}/[2\psi(t)]}- e^{-r_{0}^{2}/[2\psi(t)]} }{[2 \pi \psi(t)]^{d/2}} & .
\end{align}
An example of prediction obtained with this theory is presented on Fig.~\ref{FigTestTheory}(d).

\subsection{Rare event kinetics}
\label{RESection}

\subsubsection{General formulas}
Let us now describe how the formalism can be adapted to the study of rare event kinetics. The problem is now to determine the FPT of a non-Markovian reaction coordinate $x(t)$ to a target at $x=0$ that is rarely reached, because a large energy barrier has to be crossed. 
Obviously, equation (\ref{GenMFPT}) is still valid, and we  stress the following key points:  (i) first, as long as $x_0$ is not in the close vicinity of the target, $p(0,t)$ is  exponentially small (with noise intensity) at all times, (ii) second, the probability $p_\pi(x,t)$ to revisit the target after a time $t$ is exponentially small at long times, but finite at times that immediately follow a FPT event, when $x$ is still close to the target. This suggests that the integral (\ref{GenMFPT}) is dominated by this short time contribution, where $p_\pi$ can be replaced by its value obtained by considering the (Gaussian) linearized dynamics around the target point. Let us assume that the trajectories of $x(t)$, conditioned to $x(0)=0$, are Gaussian with covariance $\sigma$ given by (\ref{sigmaEq}) and average $m_s(t)$. Then, we approximate the dynamics after the first passage by a Gaussian dynamics, with same covariance $\sigma$, and mean $\mu(t)$. With these approximations, the estimate (\ref{GenMFPT}) becomes \cite{levernier2020kinetics}
\begin{align}
 \langle T \rangle p_s(0)\simeq\int_0^\infty dt \  e^{-\mu(t)^2/[2\psi(t)]}/[2\pi\psi(t)]^{1/2} \label{EqTS},
\end{align} 
and the self-consistent equation (\ref{05942}) for $\mu(t)$ becomes  \cite{levernier2020kinetics}
\begin{align}
&\int_0^\infty  dt \ \frac{e^{- \mu(t)^2/(2\psi(t))}}{\psi(t)^{1/2}} \Bigg\{\mu(t+\tau)  -\mu(t)\frac{\psi(t+\tau)+\psi(t)-\psi(\tau)}{2\psi(t)}-m_s(\tau)\Bigg\}=0. \label{EqMuG}
\end{align}
These equations are obtained from Eqs.~ (\ref{GenMFPT}) and (\ref{05942}) by neglecting the terms proportional to $p(0,t)$, since, according to the hypothesis (ii), this propagator is exponentially small at all times.  The above equation suggests a two-step strategy to obtain $\langle T \rangle$. The first step consists in characterizing the static quantity $p_s(0)$ ; for equilibrium systems one obtains  $p_s(0)\propto e^{-E/k_BT}$ and in particular  $\langle T\rangle$ follows an Arrhenius-like law. The second step consists in analyzing the dynamics of $x(t)$ in the vicinity of the target to deduce $\mu(t)$. We note that this theory holds for general  equilibrium and non-equilibrium systems  as well. For equilibrium systems, for which one can write a Generalized Langevin equation linearized around the target point, one has
\begin{align}
&\int_0^t dt' \dot{x}(t')K(t-t')= F + \xi(t), & \langle \xi(t)\xi(t')\rangle=k_BT K(\vert t-t'\vert),
\end{align}
where $F=-\partial_z U(z)\vert_{z=0}$ is the opposite of the slope of the potential at the target and $K$ is a memory friction kernel. Analyzing this equation leads to the following form of the fluctuation-dissipation theorem
\begin{align}
m_s(t)=\frac{  F  \psi(t)}{2k_BT}. 
\end{align}
Inserting this expression in Eqs.~(\ref{EqMuG}) and (\ref{EqTS}) leads to the determination of $\langle T\rangle$ as a function of  the local slope $F$ of the potential  near the target, the thermal energy $k_B T$, the MSD characterizing the dynamics $\psi(t)$, and  the equilibrium probability density $p_s(0)$  at the target.   As in all rare event theories the initial state does not matter since it is forgotten much faster than the time to reach the target. For the Markovian (diffusive) case with $\psi(t)\propto t$, there is an obvious solution $\mu(t)= m_s(t)$. For non-Markovian variables, this relation does not hold and the  future trajectory $\mu(t)$ reflects the state of the non-reactive degrees of freedom at the FPT.  

\subsubsection{Scale invariant processes and link with Pickands' constants}
In the case of a MSD with  power-law scaling at short times, with $\psi(t)=\kappa t^{2H}$, where $0<H<1$,  Eq.~(\ref{EqMuG}) predicts that $\mu$ takes the scaling form
\begin{align}
&\mu(t)= \frac{k_BT}{F}f\left(   t /t^*  \right), &t^*= \left( \frac{k_BT}{\sqrt{\kappa}\  \vert F\vert}\right)^{1/H} \label{DeftStar},
\end{align}
and the mean FPT reads
 \begin{align}
& \langle T \rangle   p_s(0) = \frac{A_H (k_BT)^{\frac{1-H}{H}}}{  \vert F\vert ^{  \frac{1-H}{H}} \kappa^{\frac{1}{2H}}}, & A_H=\int_0^{\infty}du \ \frac{e^{-\frac{f^2(u)}{2u^{2H}}}}{\sqrt{2\pi} u^{H}}, \label{ResultScaling} 
\end{align}  
where $f$ satisfies 
\begin{align}
&\int_0^\infty  dt \ \frac{e^{- f(t)^2/(2t^{2H})}}{t^H} \Bigg\{f(t+\tau)  -f(t)\frac{(t+\tau)^{2H}+t^{2H}-\tau^{2H}}{2t^{2H}}+\frac{\tau^{2H}}{2}\Bigg\}=0. \label{Eq_f}
\end{align}
The formula (\ref{ResultScaling}) provides an explicit asymptotic relation for the mean FPT, as a function of the subdiffusion coefficient $\kappa$, the local force $F$ and the  temperature $k_BT$, and $A_H$  depends only on $H$. 
This result is compatible with  the  mathematical results of Pickands \cite{pickands1969asymptotic} obtained when $x(t)$ is a Gaussian process (whereas here we only assume that $x(t)$ is locally Gaussian, in the vicinity of the target), if one identifies
\begin{align}
&A_H=2^{1/\alpha}\mathcal{H}_\alpha, &\alpha=2H,
\end{align}
where $\mathcal{H}_\alpha$ are Pickands' constants. Our theory therefore provides approximations of $\mathcal{H}_\alpha$. Of note, the formalism presented here can be analyzed in the limit $\varepsilon=H-1/2\to0$, leading to 
\begin{align}
A_H=2+4\varepsilon(\gamma_e -\ln2) +\mathcal{O}(\varepsilon^2),
\end{align}
where $\gamma_e$ is Euler's constant. This result agrees with the calculation of Ref.~\cite{delorme2017pickands}.

\subsubsection{Example of flexible chain reaching a long extension}

As an example of application of our formalism, we consider the ``simple'' problem of determining the mean time  for a flexible chain to reach  a given large extension. This problem plays a key role in the quantification of the mechanisms of  constraint release  that determine the rheological properties of untangled star polymers ~\cite{milner1997parameter,milner1998reptation}, and is also involved in ligand adhesion kinetics when mediated by flexible linkers~\cite{jeppesen2001impact}; it has been reconsidered in Ref.~\cite{cao2015large}. We assume here  that the polymer chain can be described by  an harmonic chain of $N$ phantom beads with friction drag $\gamma$ linked by springs of stiffness $k$, whose dynamics satisfies
\begin{align}
& \dot x_i=\tau_0^{-1}(x_{i+1}-2x_i+x_{i-1})+f_i(t) \label{EqRouse}, &\langle f_i(t)f_j(t')\rangle=2\frac{l_0^2}{\tau_0}  \delta(t-t') \delta_{ij},
\end{align}  
where $l_0=\sqrt{k_BT/k}$ is the typical bond length, and $\tau_0=\gamma/k$ the typical relaxation time of a single bond. The first monomer is fixed, $x_1=0$, and we study the mean time $\langle T \rangle$ for the second polymer end $x(t)=x_N(t)$ to reach a threshold value  $x=z$. The energy at fixed $z$  is given by $ E=k z^2/(2N)$, and we assume $E\gg k_BT$,  so that first-passage  events to $z$ are rare. 

 \begin{figure}%
  \centering
\includegraphics[width=\linewidth]{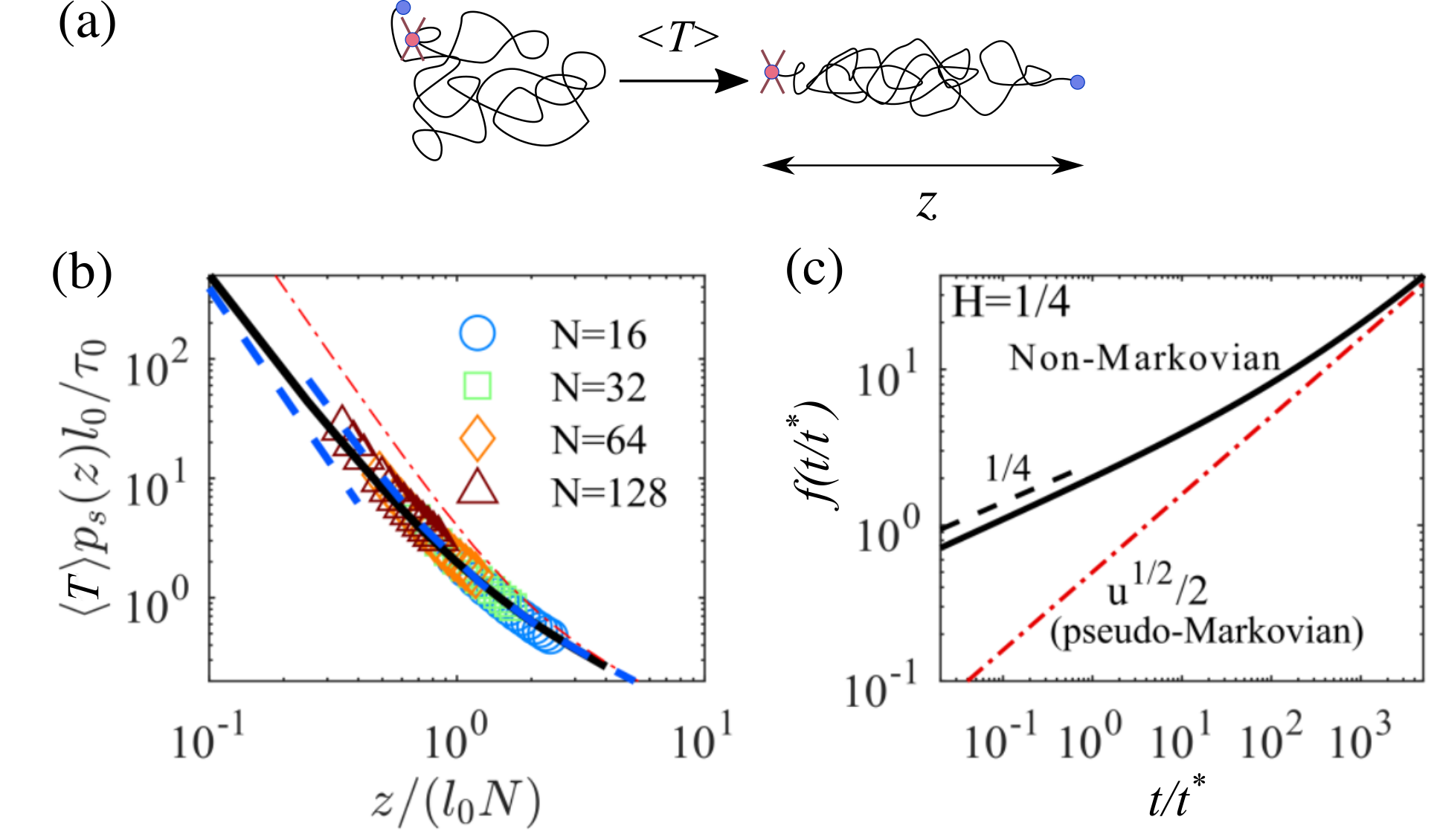}
\caption{(a) Sketch of the problem: considering an attached flexible polymer,  we compute the first time that a large extension $z$ is reached.  
(b) Scaled mean FPT as a function of the  rescaled variables $z/(l_0N)$. Symbols: simulation data of Ref.~\cite{cao2015large}. Black line: non-Markovian theory. Dashed blue line: asymptotic results (\ref{ResultNM}) of the theory. Red semi-dashed line: pseudo-Markovian approximation, with $\mu\simeq m_s$. (c) Rescaled average trajectory after the FPT $f(t/t^*)$ [see Eq.~(\ref{DeftStar})] obtained by solving numerically (\ref{Eq_f})  for  $H=1/4$. The dashed red line would be the future trajectory by assuming equilibrium at initial time. Adapted from Ref.~\cite{levernier2020kinetics}.}
\label{figCompCao}%
\end{figure}

Using the formalism above, one readily finds that the rescaled mean FPT $\langle T\rangle p_s(z)$ can be expressed as a function of the parameter $z/N$. This scaling is checked by comparing with the data of Ref.~\cite{cao2015large} on figure \ref{figCompCao}(b), and agrees well with the numerical solution of the above  equations. Furthermore, one can derive explicit asymptotic laws~\cite{levernier2020kinetics}
\begin{align}
\langle T \rangle p_s(z) \simeq 
\frac{\tau_0}{l_0 }\begin{cases}
0.39  (Nl_0/z)^3 & (l_0\sqrt{N}\ll z\ll N l_0)\\
& \\
\frac{Nl_0}{z}\left[1+ \left(\frac{Nl_0}{z}\right)^2 \right] & (N l_0\ll z ) 
\end{cases}
\label{ResultNM}.
\end{align}
For this problem, a general solution in the weak noise limit (small temperature limit at fixed $N$) can be found  [formula  (10.117) of Ref.~\cite{schuss2009theory}].  This result however  gives only the leading order term when  $N l_0\ll z$, meaning that this exact weak-noise approach does not capture the impact of memory effects in all the regimes. The result for very long chains in the first line of  (\ref{ResultNM}) is obtained by making use of the fact that the monomer motion is subdiffusive at short times, with $H=1/4$. The prefactor  is actually found to be eight times larger than predicted by the pseudo-Markovian approximation, indicating strong memory effects. The importance of these memory effects can be seen directly from the reactive trajectories $\mu(t)$   in  Fig \ref{figCompCao}(c) where one compares $\mu(t)$ and $m_s(t)$ in the regime of infinitely long chains: one clearly sees that $\mu(t)$ at short times displays a different scaling with $t$ as compared to $m_s(t)$, meaning that \textit{the folding dynamics after a first passage [$\mu(t)]$ is infinitely slower than the folding dynamics $m_s(t)$ starting from an equilibrium state} (where the chain is conditioned to be stretched). This means that the return probability $p_\pi(z,t)$ is largely overestimated by the pseudo-Markovian approximation.  
 
\section{Survival probability of unconfined compact non-Markovian random walks}
\subsection{Stochastic Processes with stationary increments}
\label{SurvProbSection}

Let us now focus on the case of an unconfined random walk, for which one cannot define a stationary PDF. Here, we focus on non-smooth Gaussian symmetric random walks with stationary increments, with MSD $\psi(t)\simeq\kappa t^{2H}$ at long times, with $0<H<1$. In this case, for compact random walks (satisfying $dH<1$), it is well known that the mean time to reach a target is infinite, and the first passage statistics is characterized by the probability of not having reached the target (survival probability, $S(t)$) at long times, 
\begin{align}
S(t)\underset{t\to\infty}{\simeq} S_0/t^\theta, \label{DecayS}
\end{align}
where $\theta$ is known as a  \textit{persistence exponent} and $S_0$ is a prefactor. In one dimension, $\theta=1-H$ can be obtained  \cite{ding1995distribution,Krug1997,Molchan1999}, and its generalization to $d$-dimension is $\theta=1-dH$ (see Ref.~ \cite{levernier2018universal} and below). This law is valid for compact processes, for which the probability to reach a target, even  pointlike, in infinite space is equal to one. 

At long times, the PDF of the position $p(\ve[r],t)$ decays as a power-law,
\begin{align}
p(\ve[0],t)\underset{t\to\infty}{\simeq} K/t^{d H}\label{Def_K},
\end{align}
where, in the case of  Gaussian processes, $K=1/[2\pi\kappa]^{d/2}$. Let us briefly describe how to obtain the prefactor $S_0$  in (\ref{DecayS}). We start with the generalized renewal equation (\ref{Renewal}) (adapted to the $d$-dimensional case), which we integrate over $t\in]0,A[$ to obtain
\begin{align}
\int_0^A dt p(\ve[0],t)=\int_0^A dt\int_0^t d\tau F(\tau)p(\ve[0],t\vert\text{FPT}=\tau)\nonumber\\
=\int_0^A du \int_0^{A-u} d\tau F(\tau)p(\ve[0],\tau+u\vert \text{FPT}=\tau),
\end{align}
where we have used $t=\tau+u$. As before, we consider the PDF of the position $p_\pi(\ve[x],t)$ at a time $t$ after the first passage event, given by Eq.~ (\ref{Def2PointsPPi}) (generalized to higher spatial dimensions), so that for any $A>0$
\begin{align}
\int_0^A dt [p_\pi(\ve[0],t)-p(\ve[0],t)]=\int_0^A dt \int_{A-t}^\infty d\tau F(\tau)p(\ve[0],\tau+t\vert \text{FPT}=\tau). \label{0421}
\end{align}
The right hand side can be estimated by assuming that, at long times, one has the following decoupling approximation:
\begin{align}
 p( \ve[0], (u+v)A\vert \text{FPT}=uA) \underset{A\to\infty}{\simeq} p_\pi(\ve[0],vA)\sim K/(vA)^{dH}. \label{SuppDecApprox}
\end{align}
Using $t=uA$ and $\tau=vA$ in  (\ref{0421}) and the above approximation, we obtain
\begin{align}
\int_0^A dt [p_\pi(\ve[0],t)-p(\ve[0],t)]=A^{1-dH-\theta} \int_0^1 du \int_{1-u}^\infty dv \frac{KS_0\theta}{u^{\theta+1}v^{dH}}\label{4321}.
\end{align}
Now, for $A\to\infty$, we see that the left-hand side is nothing but the value of the mean FPT for large volumes, rescaled by the volume. Since its value is finite for large $A$, we deduce that the right-hand side of the above equation must also reach a finite value for large $A$. This leads to \cite{levernier2019survival}
\begin{align}
&\theta=1-dH, & S_0 \simeq \frac{\sin (\pi d H) }{\pi K } \times \lim_{V\to\infty} \frac{\langle T\rangle}{V} \label{prefactor},
\end{align}
where $K$ is defined in Eq.~(\ref{Def_K}). For non-Markovian processes, this result requires the ``long time decoupling approximation'' (\ref{SuppDecApprox}). However, relaxing this approximation (by introducing a scaling function $G(u/v)$ in (\ref{SuppDecApprox})) would only change the results by a scaling factor, while the dependence of $S_0$ on the initial distance to the target  would be taken into account in $\langle T\rangle/V$. 

\begin{figure}[h!]
\centering
\includegraphics[width=11cm,clip]{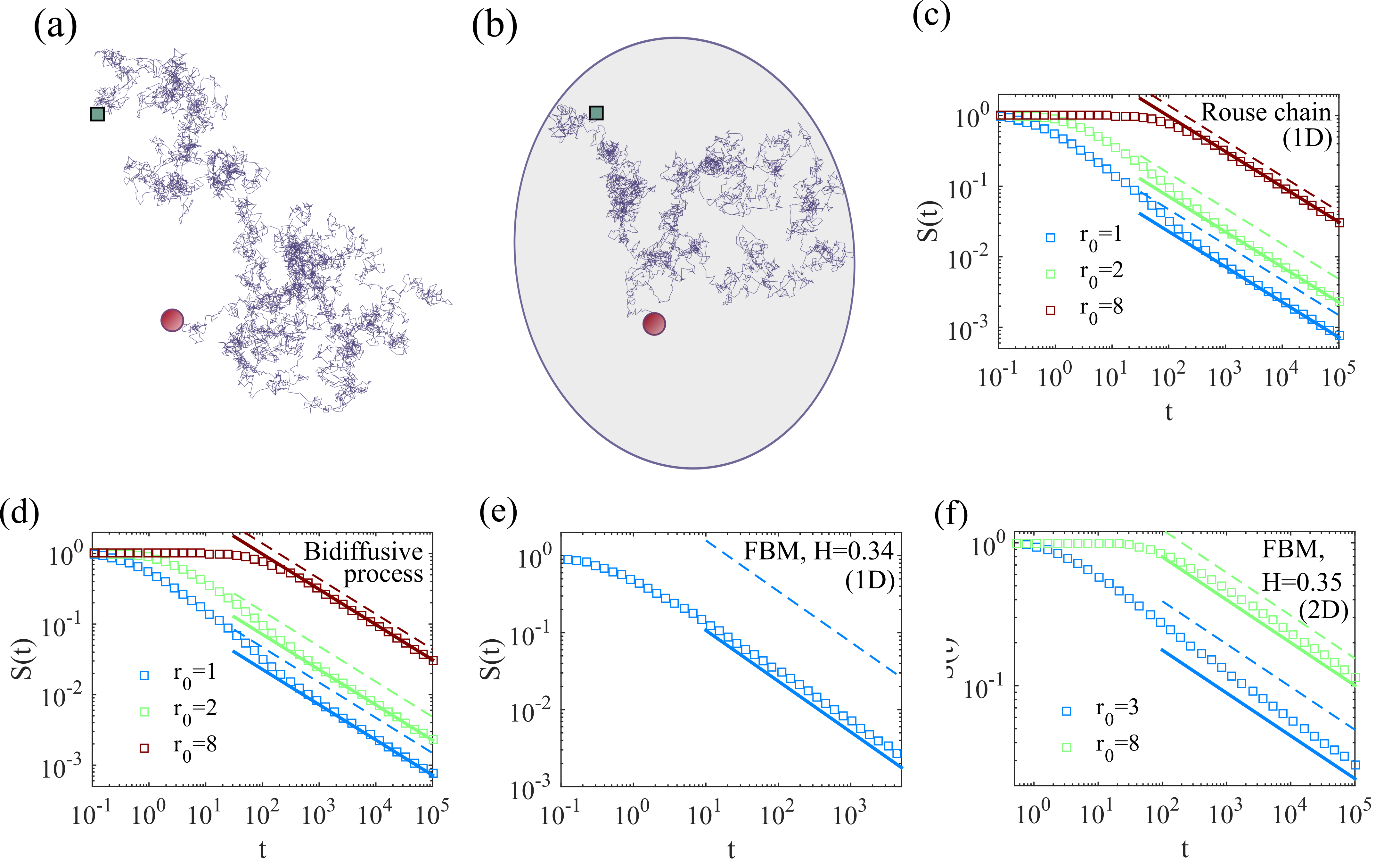}   
\caption{(a),(b) Sketch of  two first passage problems, without (a) and with (b) confinement for the same stochastic process.  In case (a)  the mean FPT to reach the target is infinite, but it is finite in case (b) and can be used to characterize the long-time decay of the survival probability in problem (a). (c-f) figures showing simulation results for $S(t)$ compared to the prediction obtained with (\ref{prefactor}) and the present formalism, for various stochastic processes indicated in each sub-figure. Adapted from Ref.~\cite{levernier2019survival}. }
\label{FigSurvival}
\end{figure}  

Equation  (\ref{prefactor}) indicates that the prefactor $S_0$ is directly proportional to the mean FPT for the same random walker in confinement, thus enabling one to calculate $S_0$ by the same methods that were used in confinement.  Figure \ref{FigSurvival} compares our predictions for $S_0$ with simulations of various processes.  In fact, Eq.~(\ref{prefactor}) is very general: it holds for compact processes with and without memory, for any spatial dimension (including fractal spaces), for perfect or imperfect reactivity...  We could also use this formula for other forms of confinement, as provided by considering the resetting of the random walker with some resetting rate. The above formula   (\ref{prefactor}) would then link the prefactor of the survival probability in absence of resetting to the mean FPT to the target in presence of resetting for low resetting rates, provided that one replaces $1/V$ by the stationary probability to be at the target in presence of resetting. This can be checked directly by using the formulas present in Ref.~\cite{pal2017first}. 

Last, we note that the above equation gives an information on the first passage time density even in confinement at intermediate times when the boundaries are not reached. Hence, although the mean FPT does not characterize alone the full distribution (which is not exponential), the above formula indicates that the mean FPT provides valuable informations on the shape of the FPT density. 

\subsection{Non-stationary initial conditions: Survival probability after a quench }
\label{PersistenceSec}

In this last section, we  discuss the case  of non-stationary initial conditions. This could typically occur for example after a sharp change of  temperature of the whole system at initial time (a quench). For such non-stationary initial conditions, a classical problem is that of the determination of persistence exponents $\theta$ \cite{ReviewBray}, which can be non-trivial  even in simple models, with rare   exact results \cite{poplavskyi2018exact,derrida1995exact,dornic2018universal}. Here, we focus on Gaussian processes that are transiently aging: we assume that the PDF of $x(t+t_1)$, conditioned to the value of $x(t)$, reaches  for large $t$ a finite steady state value that does not depend on $t$.  An example of such processes is provided by the dynamics of polymers or interfaces exposed to a temperature quench at $t=0$, in which case the covariance takes the form
\begin{align}
\sigma(t,t')\propto T_0(t^{2H}+t'^{2H})+(1-T_0)(t+t')^{2H}-\vert t-t'\vert^{2H},\label{CovInt}
\end{align} 
where $T_0$ is the temperature at past times $t<0$ (not to be confounded with the first passage time $T$), while the temperature of the dynamics is taken as unity for $t>0$. Of note, the temperature $T_0$ before the quench can be lower or larger than the temperature $T_0=1$ of the dynamics for $t>0$. This covariance function (\ref{CovInt}) is relevant for the dynamics of several models of interfaces and polymer dynamics \cite{Krug1997}. For example $H=1/4$ corresponds to a Rouse chain and to Edwards-Wilkinson interface, $H=3/8$ to the Mullins-Herring interface (or semiflexible chains), other values of $H$ can be obtained for macromolecules with fractal (hyperbranched) architecture  \cite{dolgushev2015contact}.   

There exists a fairly general method to study persistence exponents, which consists in defining a stationary process $X(\tau)\equiv  x(e^{\tau})/\langle x(e^{\tau})^2\rangle^{1/2}$ (where $\tau=\ln t$), and analyzing the first passage properties of $X$ within the independent interval approximation. This method applies to processes that are aging at all times, such as the random acceleration process \cite{BURKHARDT1993,deSmedt2001partial} or  systems in which the dynamics occurs at zero temperature \cite{poplavskyi2018exact,dornic2018universal,derrida1994non,derrida1995exact,Majumdar1996,Majumdar1996uq,Derrida1996,watson1996persistence,newman2001critical}. However, it appears that this method is not applicable to transiently aging processes such as those described by the covariance (\ref{CovInt}). Technically speaking, the reason of this failure is that the obtained process $X(\tau)$ is not smooth, so that the independent interval approximation cannot be applied to analyze its first passage properties. The only available methods to study persistence exponents are perturbative methods (around Markovian processes) and the derivation of upper and lower bounds \cite{Krug1997}. 

Here, we may establish a link between the long time behavior of trajectories after a first passage event and the persistence exponents. Indeed, assuming that $\theta<1-dH$, we may use Eq.~(\ref{4321}) (which is still valid here) and see that it can be satisfied only if
\begin{align}
p_\pi(\ve[0],t)-p(\ve[0],t)\underset{t\to\infty}{\sim} \frac{1}{t^{dH+\theta}}\label{04391}.
\end{align} 
Here, we focus on processes with zero mean (in practice, this is realized by assuming that  $x(t=0)=0$, and considering first-passage events only for $t>\Delta$ for a fixed $\Delta>0$). Hence, we  assume that the trajectories after a first passage event are Gaussian, with zero mean, and covariance $\sigma_\pi(t,t')$ (for each spatial coordinate). In this case, $p_\pi(\ve[0],t)\simeq 1/[2\pi \sigma_\pi(t,t)]^{d/2}$, so that
Eq.~(\ref{04391}) can be satisfied only if 
\begin{align}
\sigma_\pi(t,t) -\sigma(t,t)\underset{t\to\infty}{\sim} t^{2H-\theta} \label{LongTimeSigmaPi}.
\end{align}
This means that \textit{the calculation of the exponent $\theta$ amounts to that of the covariance $\sigma_{\pi}(t,t')$ of the trajectories in the late future after a first-passage event}. 

\begin{figure}[h!]
\centering
\includegraphics[width=0.95\linewidth,clip]{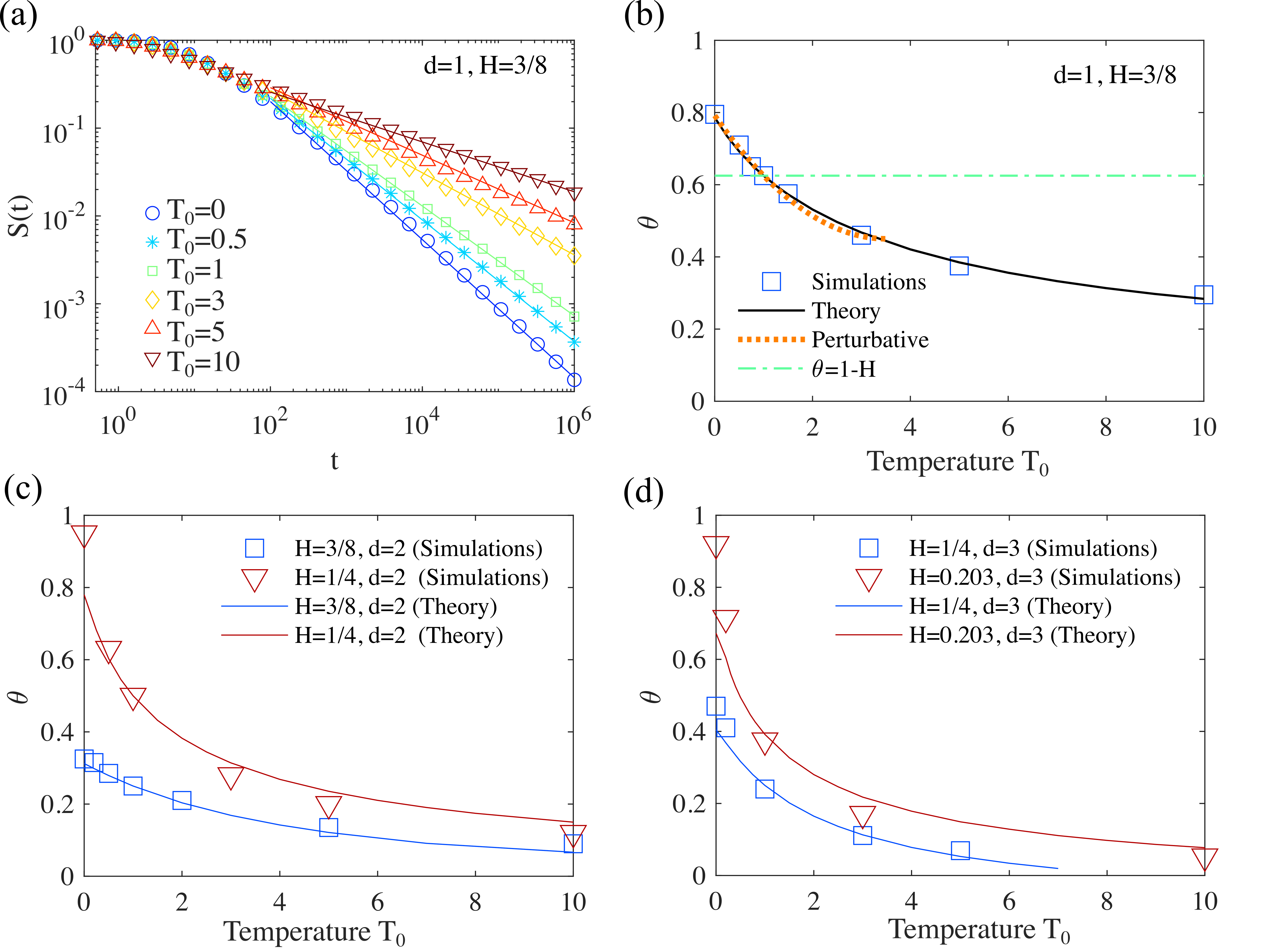}   
\caption{ Persistence for ``quenched fBM'', whose covariance follows Eq.~(\ref{CovInt}). (a) Example of survival probabilities for interface dynamics with $H=3/8$ at different temperatures. The slopes of the continuous lines is the value of $\theta$ predicted by our approach. (b),(c),(d): Systematic comparison of $\theta$ as measured in simulations versus theoretical values for different $T_0$ and (b)  $d=1$, (c) $d=2$ and (d) $d=3$ spatial dimensions. In (b), the perturbation results are Eq.~(\ref{PertResQuenchedfBM}). 
 Adapted from Ref.~\cite{Levernier2022Everlasting}}
\label{FigPersistence}
\end{figure}

A procedure to find the persistence exponent  then consists in writing a self-consistent equation for $\sigma_\pi$ (as was done to determine the average trajectories $\mu(t)$), followed by an analysis of this equation at long times to identify $\theta$ via Eq. (\ref{LongTimeSigmaPi}). This procedure is done in Ref. \cite{Levernier2022Everlasting} and leads to a linear integral equation of the form 
\begin{equation}
\label{eqint}
\int_0^1 K_\theta(u,v)\left[z_\theta(u)-z_\theta(1)\left(1-\frac{(2H-\theta) (1-u)}{2}\right)\right] {\rm d}u=f_\theta(v),
\end{equation}
where $\sigma_\pi(t,t')-\sigma (t,t')\simeq t^{2H-\theta}z(t/t')$ at long times, and $K_\theta,f_\theta$ can be calculated in terms of $\sigma$. Finding the value of $\theta$ for which $z$ does not diverge then leads to the prediction of the value of $\theta$. This leads to the results presented in Fig.~\ref{FigPersistence}. There, one clearly sees that $\theta$ depends on the temperature $T_0$. It is found that the theory captures quantitatively the dependence  of the persistence exponents on the temperature quench for all these models. Of note,  perturbative analysis  of the problem for small values of $\varepsilon=H-1/2$ can be conducted and yields
\begin{align}
&\theta= 1-H-2(\sqrt{2}-1)(1-T_0)\varepsilon +\{a_1 (1-T_0)[a_2 +(1-T_0)] \}\varepsilon^2+\mathcal{O}\left(\varepsilon^3\right), \label{PertResQuenchedfBM}
\end{align}
where analytical expressions of $a_1$ and $a_2$ can be obtained, with numerical estimates $a_1\simeq1.77$, $a_2\simeq1.28$. Interestingly, in the particular cases $T_0=0$ and $T_0=1$, the first order terms coincide with the exact first order solution of  Ref.~\cite{Krug1997}, which points towards the exactness of this approach at this order.

The dependence on the temperature quench shows that the exponents that quantify the kinetics of first passage to a  target are  significantly modified if  the system is prepared with non-stationary initial conditions. Of note, this result holds in space dimensions $d\ge1$,  and  in particular for $d=3$. In the context of chemical reactions, this means that the kinetics  of  reaction involving complex macromolecules could be significantly modified by a change of initial conditions, obtained either by   a temperature quench or by imposing a constraint, such as a geometric confinement or an external field, that would be relaxed at $t=0$. The above method  then allows to quantify the kinetics of such reactions.

\section{Conclusion}

As a summary, this chapter has reviewed an approach that allows for a quantitative determination of  the first passage properties of non-Markovian random walkers to target sites in various geometries and space dimensions. The originality of the approach is that it relies on the analysis of the trajectories of the random walker after the first passage event, as documented in references~\cite{guerin2016mean,Levernier2022Everlasting,levernier2020kinetics,levernier2019survival}; this has allowed in recent years for  new developments in paradigmatic examples of target search problems. This approach was first applied to prototypical problems of target search within confined spaces; it was  next extended to the case of rare events kinetics, and finally to the case of infinite geometries, with the refined determination of  the survival probability and persistence of processes featuring stationary increments or transient aging. 
A key point to assess the importance of memory effects lies in the fact that for non-Markovian dynamics,  trajectories after the first passage event differ from those originating from an equilibrium state. This discrepancy comes from the fact that the degrees of freedom of the bath are not in an equilibrium state at the instant of the first passage, as reported  in specific examples  in Refs.~\cite{Guerin2012a,dolgushev2015contact,Levernier2015,benichou2015mean,Guerin2013,Guerin2013a}. The approach that we have presented shows that one does not need to describe explicitly the distribution of the degrees of freedom of the bath; the position of the random walker is in principle sufficient. Interestingly, this approach agrees with the available exact results obtained perturbatively for  examples of weakly non-Markovian processes. Of note, similar arguments were used recently to predict the shape of the distribution of first passage times  \cite{sakamoto2023first}.  Finally, these series of studies reveal that the characterization of the non-equilibrium state of the system at the instant of first passage is key to derive first-passage kinetics, and they provide a new methodology, via the analysis of trajectories after the first-passage, to make it quantitative.


\end{document}